\documentclass[preprint,preprintnumbers,amsmath,amssymb]{revtex4}
\usepackage{graphicx}
\usepackage{subcaption}
\usepackage{chngcntr}
\usepackage{bm}
\usepackage{epsfig}
\begin{document}

\title  {Electromagnetic transitions of $(b{\bar c})$ bound system }
\author {Sonali Patnaik$^{1}$, P. C. Dash$^{1}$, Susmita Kar$^{2}$\footnote{email address:skar09.sk@gmail.com}, 
N. Barik$^{3}$}

\affiliation{$^{1}$ Department of Physics,
Siksha 'O' Anusandhan Deemed to be University, Bhubaneswar-751030, India\\
$^{2}$ Department of Physics, North Orissa University, Baripada-757003, India\\ 
$^{3}$ Department of Physics, Utkal University, Bhubaneswar-751004, India}

\begin{abstract}
We study electromagnetic transitions: $B_c^*(ns)\to B_c(ns) e^+ e^-$, $B_c^*(ns)\to B_c(n^{\prime}s) e^+ e^-$ and $B_c(ns)\to B^*_c(n^{\prime}s) e^+ e^-$ in the relativistic independent quark (RIQ) model based on a flavor-independent potential in the scalar-vector harmonic form. The transition form factors for energetically possible transitions involving $B_c$ - and $B_c^*$- mesons in ground as well as orbitally excited states are predicted in their respective kinematic range. Our predictions on decay width for the allowed and hindered transitions are found compatible with those of the model calculations based on Bethe-Salpeter approach. Predictions in this sector would not only provide more information about members of the $B_c$-family including mass splitting between vector mesons and corresponding pseudoscalar counterparts but give hints for experimental determination of unknown masses of other excited $B_c$ - and ground state of $B_c^*$-meson, which is expected at LHCb and $Z^0$ factory in near future.
\end{abstract}
	
\maketitle

\section{Introduction}
Ever since its discovery at Fermilab by CDF Collaboration \cite {A1}, $B_c$-meson has been attracting a great deal of attention both theoretically and experimentally. The mesons in the 
$b {\bar c}$  ($B_c$) family lie intermediate in mass and size between charmonium $(c \bar c)$ and bottomonium $(b \bar b)$ family, where the heavy quark interactions are believed to be understood rather well. $B_c$-meson with explicitly two heavy quarks has not yet been thoroughly studied because of insufficient data available in this sector. Even though the ground state $B_c$ of $J^P=0^-$, was found several years ago its partner $B_c^*$ of $J^P=1^-$ has not yet been seen. Earlier attempts \cite{A2,A3,A4}, to observe $B_c$ at $e^+e^-$ collider could not succeed since the luminosity and collision energy as that of LEP-I and II could result in only small statistics for $B_c$ events \cite{A5,A6,A7}. With the observation of $B_c$ at hadron colliders, TEVATRON \cite{A8,A9}, a detailed study of $B_c$ family members is expected at LHC where the available energy and luminosity are much higher than at TEVATRON, that should result in $B_c$- events thousand times more. The lifetime of $B_c$ has been measured \cite {A10, A11, A12, A13} using decay channels: $B_c^{\pm}\to J/\psi e^{\pm}{\bar \nu}_e$  and $B_c^{\pm}\to  J/\psi\pi^{\pm}$. At LHCb a more precise $B_c$ lifetime is obtained \cite {A14} using the decay mode: $B_c\to J/\psi\mu\nu_e{\mu}X$ where X denotes any possible additional particle in the final state. Recently the ATLAS Collaboration at LHC have detected excited $B_c$- state \cite {A15} through the channel: $B_c^{\pm}(2s) \to B_c^{\pm}(1s) \pi^+\pi^-$ by using $4.9fb^{-1}$  of 7 TeV and $19.2 fb^{-1}$  of 8 TeV pp collision data yielding $B_c(2s)$-  meson mass $\sim6842\pm4\pm5$ MeV. Here the problem encountered is that the messy QCD background of the hadron colliders  contaminating the environment makes precise measurement difficult and therefore observation of excited $B_c$-states and $B_c^*$ ground state is almost impossible at LHC. In this respect, the proposed $Z^0$ factory offers conducive environment for measurement. $Z^0$-factory, an $e^+$ $e^-$ collider, running at $Z^0$ boson pole with sufficiently higher luminosity and offering relatively cleaner background is supposed to enhance the event-accumulation rate so that other excited $B_c$ states and possibly the $B^*_c$ ground state are likely to be observed in near future.

Unlike heavy quarkonia, $B_c$-meson with explicitly two heavy quark constituents  do not annihilate to photons or gluons. The ground state $B_c$-meson can therefore decay weakly through $b \to c W^-$; ${\bar c}\to {\bar s}W^-$ or decay radiatively through   $b \to b\gamma$; ${\bar c}\to {\bar c}\gamma$ at the quark level. A possible measurement of radially excited states of $B_c$ via $B_c(ns)\to B_c\pi \pi$ at LHC and  $Z^0$ factory is discussed in Ref.\cite {A16}. However the splitting between $B_c(1s)$ and its nearest member $B_c^*(1s)$ due to possible spin - spin interaction estimated in the range $30\leq\Delta m\leq50$ MeV \cite {A17} forbids the process $B^*_c\to B_c+\pi^0(\eta,\eta^{\prime})$ by energy momentum conservation. Therefore the dominant decay modes in this sector are the magnetic dipole radiative decays of the type $B_c^*(ns)\to B_c(ns)\gamma$ and $B_c(ns)\to B^*_c(n^{\prime}s)\gamma$ with $n>n^{\prime}$. Another decay mode of interest  is  $B_c^*(ns)\to B_c(ns) e^+ e^-$ which is also governed by electromagnetic process where the emitted photon is an off shell virtual one.  Compared to radiative decays emitting real photons, the rate of these decay processes is thought to be highly suppressed due to a tight three body phase space and an extra electromagnetic vertex.  These processes are more interesting theoretically because the lepton pair ($e^+,e^-$) product could be easily caught by the detector as clear signals. Being charged particles their track can be more easily identified than that of the neutral photon emitted in M1 radiative decays of $B_c$ and $B_c^*$. 

Several theoretical attempts \cite{A17, A18, A19,A20,A21,A22,A23,A24,A25,A26,A27,A28,A29,A30,A31} including different versions of potential models based on Bethe-Salpeter(BS) approach, light front quark (LFQ) model, QCD sum rules and Lattice QCD (LQCD) etc.  have predicted the $B_c$-spectrum, its mass and decay widths. We have analyzed various M1 transitions of the type $V\to P\gamma$ and $P\to V\gamma$  in the light and heavy flavor sector within and beyond static approximation \cite{A32} in the framework of relativistic independent quark (RIQ) model. We have also studied the $q^2$-dependence of relevant transition form factors and predicted decay widths for radiative decays of heavy mesons in the charm and bottom flavor sector \cite{A33} and recently predicted the magnetic dipole radiative transitions of the ground and excited $B_c$ and $B_c^*$ mesons \cite{A34} in good comparison with other model predictions.  The applicability of RIQ model has already been  tested in describing wide ranging hadronic phenomena including the static properties of hadrons \cite{A35} and various decays such as radiative, weak radiative and rare radiative \cite{A32, A36}: leptonic and weak leptonic 
\cite{A37,A38} radiative leptonic \cite{A39}; semileptonic \cite{A40},and non-leptonic \cite{A41} decays of mesons  in the light and heavy flavor sector. KE Hong Wei $\it etal$.  \cite{A17} in their analysis of magnetic dipole transitions predicted $B_c^*(ns)\to B_c(ns)e^+e^-$ and $B_c(ns)\to B^*_c(n^{\prime}s)e^+e^-$ with $n>n^{\prime}$. We would like to extend the applicability of RIQ model to describe such decay modes  involving $B_c$ and $B_c^*$ mesons  in their ground and excited states . Such a study would be helpful in extracting more information about members of $B_c$ family, determining mass splitting and predicting the decays widths.
       
The paper is organized as follows: In Sec. II
we give a brief account of the RIQ model and describes model expressions for the transition form factors and decay width.
In Sec.-III we provide our numerical results and discussion. Section IV encompasses our summary and conclusion.

\section{Transition matrix element, transition form factor and decay width in RIQ model}

The RIQ model framework has been discussed in earlier applications of the model to a wide range of hadronic phenomena \cite{A32, A33, A34, A35, A36, A37, A38, A39, A40, A41}. For the sake of completeness  we provide here a brief description of the model framework and model expressions for constituent quark orbitals along with corresponding momentum probability amplitudes in the Appendix. In a field-theoretic description of any decay process, which in fact occurs physically in the momentum eigenstate of participating mesons, a meson state such as $|B_c(\vec P,S_V)>$ is considered at definite momentum $\vec P$ and spin state $S_V$ in terms of appropriate wave packet \cite{A32,A33,A34, A36,A37,A38,A39,A40,A41}as:
\begin{equation}
|B_c(\vec P, S_V)> = {\hat \Lambda_{B_c}(\vec P,S_V)}|(\vec p_b,\lambda_b);(\vec p_c,\lambda_c)>
\end{equation}
where, $|(\vec p_b,\lambda_b);(\vec p_c,\lambda_c)>=\hat b_b^\dagger (\vec p_b,\lambda_b) 
{\hat {\tilde{b}}}_c^\dagger (\vec p_c,\lambda_c)|0>$ is a Fockspace representation of the 
unbound quark b and antiquark $\bar c$ in a 
color-singlet configuration with their respective momentum and spin as  $(\vec p_b,\lambda_b)$ and $(\vec p_c,\lambda_c)$. 
Here $\hat b_b^\dagger (\vec p_b,\lambda_b)$ and  ${\hat {\tilde{b}}}_c^\dagger (\vec p_c,\lambda_c)$
are respectively the quark and antiquark creation operators. ${\hat \Lambda}_{B_c}(\vec P,S_V)$ 
represents a bag like integral operator taken in the form: 
\begin{equation}
{\hat \Lambda}_{B_c}(\vec P,S_V)=\frac{\sqrt 3}{\sqrt{N_{B_c}(\vec P)}}\;
\sum _{{\lambda_b},{\lambda_{\bar c}}}\zeta_{b,{\bar c}}^{B_c}(\lambda_b,\lambda_{\bar c})
\int d^3{\vec p}_{b}\;d^3{\vec p}_{\bar c}\;\delta^{(3)}(\vec p_{b}+\vec p_{\bar c}-\vec P)
{\cal G}_{B_c}(\vec p_{b},\vec p_{\bar c})
\end{equation}
Here $\sqrt 3$ is the effective color factor, 
$\zeta_{b,{\bar c}}^{B_c}(\lambda_b,\lambda_{\bar c})$
stands for appropriate SU(6)-spin flavor coefficients for the meson. $N(\vec P)$ is the meson-state normalization which can be realized from
$<{B_c}(\vec P)\mid {B_c}({\vec P}\;^{\prime})>=\delta ^{(3)}(\vec P-{\vec P}\;^{\prime})$ 
in an integral form
\begin{equation}
N(\vec P)=\int d^3{\vec p}_b\;\mid {\cal G}_{B_c}({\vec p}_b,\vec P-{\vec p}_b)\mid ^{2}
\end{equation}
Finally  $ {\cal G}_{B_c}({\vec p}_b,\vec P-{\vec p}_b)$ is the effective momentum profile function for the quark-antiquark pair which in terms of individual momentum probability amplitudes: $G_b(\vec p_b)$ and ${\tilde G}_c(\vec p_c)$ for quark b and antiquark $\bar c$ respectively, is considered in the form 
\begin{equation}
{\cal G}_{B_c}({\vec p_b},{\vec p_{\bar c}})=\sqrt{G_b(\vec p_b){\tilde G}_{\bar c}(\vec p_{\bar c})}
\end{equation}
in a straightforward extension of the ansatz of Margolis and Mendel in their bag model analysis \cite{A42}.

In the wave packet representation of meson bound state $|B_c(\vec P,S_V)>$, the bound state character  is thought to be embedded here in ${\cal G}_{B_c}({\vec p_b},{\vec p_{\bar c}})$. Any residual internal dynamics responsible for decay process such as $B_c^*\to B_ce^+e^-$ can therefore be analyzed at the level of otherwise 
free quark and antiquark using appropriate Feynman diagrams. 
Total contributions from Feynman diagram provides the constituent level S-matrix element $S_{fi}^{b \bar c}$ which when operated by the operator  ${\hat \Lambda}_{B_c}(\vec P,S_V)$ gives meson level effective S-matrix element $S_{fi}^{B_c}$ as
\begin{equation}
S_{fi}^{B_c} = {\hat \Lambda}_{B_c}(\vec P,S_V) S_{fi}^{b\bar c}
\end{equation}
The hadronic matrix element for :$B_c^*\to B_c\ e^+e^-$ finds a covariant expansion in terms of transition form factor as $F_{B_c^*B_c}(q^2)$
\begin{equation}
<B_c(k)|J^{em}_{\mu}|B_c^*(P,h)>=ie\epsilon_{\mu\nu\rho\sigma}\epsilon^{\nu}
(P,h)(P+k)^{\rho}(P-k)^{\sigma}F_{B_c^*B_c}(q^2)
\end{equation}
where, $q=(P-k)=k_1+k_2$ is the four momentum transfer, $k,k_1$, $k_2$ are four momentum of $B_c$, electron and positron, respectively and $\epsilon_{\nu}(P,h)$
is the polarization vector of $B_c^*$ with four momentum {\it P}
and helicity {\it h}. For transition $B_c^*\to B_ce^+e^-$, the kinematic range of $q^2$ is $(2m_e)^2\leq q^2\leq(m_{B^*_c}-m_{B_c})^2$. The $q^2$-dependence of the form factor can be studied using the expression for $F_{B_c^*B_c}(q^2)$ obtainable in the RIQ model.

\begin{figure}
	\begin{center}
		\includegraphics[width=12 cm,height=7cm]{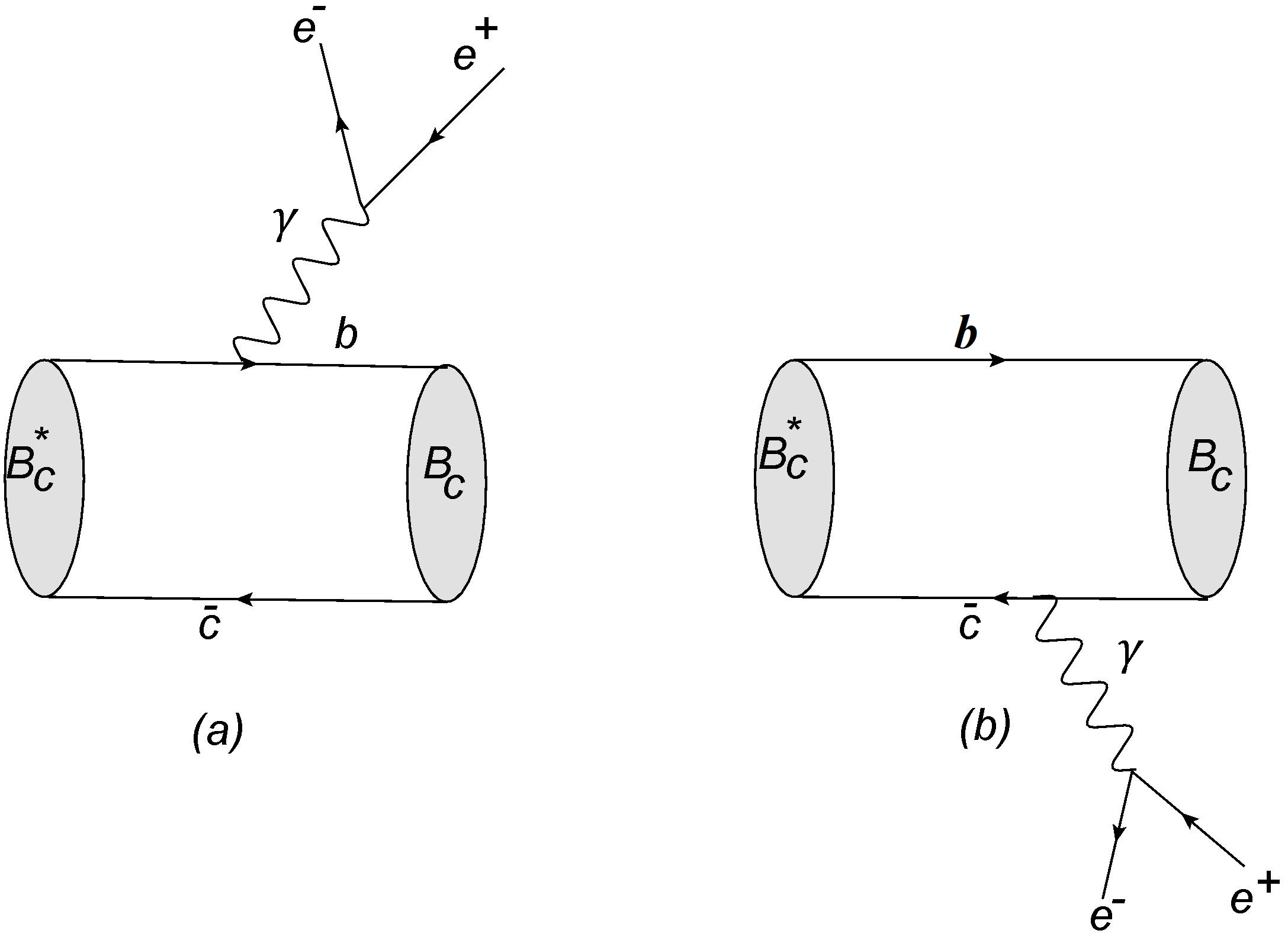}
	\end{center}
	\caption{{Lowest order Feynman diagram contributing electromagnetic transition}}
\end{figure}
The decay process $B_c^* \to B_c e^+e^-$ as depicted in Fig.1(a,b) is thought to be predominantly a double-vertex electromagnetic process governed by photon emission at the photon-hadron vertex from independently confined quark $b$ as well as antiquark $\bar c$ confined in the meson bound state $|B^*_c(\vec P,S_V)>$.
The emitted photon is an off-shell virtual one which ultimately leptonizes into pair of leptons $(e^-, e^+)$. The S-matrix element for the process in configuration space is written as:
\begin{eqnarray}
S_{fi}=<B_c(k)e^+(k_1,\delta_1)e^-(k_2,\delta_2)|(-ie)^2\int d^4x_1 d^4x_2{\bar \psi}^{(-)}_{e^-}(x_2)\gamma^\mu{\psi}^{(-)}_{e^+}(x_2)\;
{\cal D}_{\mu\nu}(x_2-x_1)\nonumber\\
\times\sum_qe_q{\bar \psi}^{(+)}_q(x_1)\gamma^\nu{\psi}^{(+)}(x_1)
|B^*_c(\vec P,S_V)>
\end{eqnarray}
where ${\cal D}_{\mu\nu}(x_2-x_1)$ is photon propagator. Now using usual expression for photon propagator, quark and lepton field expansion and then simplifying hadronic and leptonic part separately by adopting the vacuum insertion technique, $S_{fi}$ in the $B_c^*$ rest frame is obtained in the standard form as
\begin{equation}
S_{fi}=(2\pi)^4\delta^{(4)}(k+k_1+k_2-{\hat O}M_{B_c^*})\frac{({-i\cal M}_{fi})}{\sqrt{(2\pi)^32M_{B^*_c}}}
\prod_f\frac{1}{\sqrt{(2\pi)^32E_f}}
\end{equation} 
where, the hadronic part $h_{\mu}$ is found to be
\begin{eqnarray}
h_{\mu}=\sqrt{2M_{M_c^*}2E_k}\;[e_b\int d{\vec p}_b
\frac{{\cal G}_{B^*_c}({\vec p_b},-{\vec p_b})
{\cal G}_{B_c}({\vec k}+{\vec p_b},-{\vec p_b})}
{\sqrt{2E_{p_b}2E_{p_b+k}N(0)N(\vec k)}}
{\cal C}^{B^*_cB_c}_{\lambda_b\lambda_c\lambda_b^{\prime}}\nonumber\\
-e_c\int d{\vec p}_c
\frac{{\cal G}_{B^*_c}(-{\vec p_c},{\vec p_c})
{\cal G}_{B_c}(-{\vec p_c},{\vec k}+{\vec p_c})}
{\sqrt{2E_{p_c}2E_{p_c+k}N(0)N(\vec k)}}
{\cal C}^{B^*_cB_c}_{\lambda_b\lambda_c\lambda_c^{\prime}}]
\end{eqnarray}
with
\begin{eqnarray}
{\cal C}^{B^*_cB_c}_{\lambda_b\lambda_c\lambda_b^{\prime}}
=\sum_{\lambda_b\lambda_c\lambda_b^{\prime}}\zeta^{B_c^*}_{b,c}(\lambda_b,\lambda_c)
\zeta^{B_c}_{b^{\prime},c}(\lambda_b^{\prime},\lambda_c)
{\bar U}_{b^\prime}({\vec k}+{\vec p_b},\lambda_b^{\prime})
\gamma_{\mu}U_b({\vec p}_b,\lambda_b)\nonumber\\
{\cal C}^{B^*_cB_c}_{\lambda_b\lambda_c\lambda_c^{\prime}}
=\sum_{\lambda_b\lambda_c\lambda_b^{\prime}}\zeta^{B_c^*}_{b,c}(\lambda_b,\lambda_c)
\zeta^{B_c}_{b,c^{\prime}}(\lambda_b,\lambda_c^{\prime})
{\bar V}_c({\vec p_c},\lambda_c)
\gamma_{\mu}V_c({\vec k}+{\vec p}_c,\lambda_c^{\prime})
\end{eqnarray}
and the leptonic part $l^{\mu}$ is
\begin{equation}
l^{\mu}(k_1,k_2,\delta_1,\delta_2)={\bar U}_{e^-}(k_2,\delta_2)\gamma^{\mu}V_{e^+}(k_1,\delta_1)
\end{equation}
Here the timelike component of $h_{\mu}$ in Eq.(9) vanishes identically for each combination of $B_c^*$ spin state with the singlet state of $B_c$. As a result ${\cal M}_{fi}$ is effectively expressed in terms of spacelike parts of the 
hadronic and leptonic part in the form:
\begin{equation}
{\cal M}_{fi}=e^2h_il^i(k_1,k_2,\delta_1,\delta_2)/(k_1+k_2)^2
\end{equation}
Using usual spin algebra, the non vanishing spacelike hadronic part $h_i$ is obtained as
\begin{equation}
h_i=(e_bI_b+e_cI_c)({\vec \epsilon}\times {\vec k})_i 
\end{equation}
with
\begin{eqnarray}
I_b=\sqrt{2M_{B_c^*}2E_k} 
\int d{\vec p_b}\frac{{\cal G}_{B_c^*}({\vec p_b},
-{\vec p_b}){\cal G}_{B_c}({\vec p_b}+{\vec k}, -{\vec p_b})}
{\sqrt{{2E_{p_b}2E_{p_b+k}{\bar N}_{B_c^*}(0){\bar N}_{B_c}(\vec k)}}}
\sqrt{{(E_{p_b}+m_b)\over{(E_{p_b+k}+m_b)}}}\nonumber\\
I_c=\sqrt{2M_{B_c^*}2E_k} 
\int d{\vec p_c}\frac{{\cal G}_{B_c^*}(-{\vec p_c},
	{\vec p_c}){\cal G}_{B_c}(-{\vec p_c}, {\vec p_c}+{\vec k})}
{\sqrt{{2E_{p_c}2E_{p_c+k}{\bar N}_{B_c^*}(0){\bar N}_{B_c}(\vec k)}}}
\sqrt{{(E_{p_c}+m_c)\over{(E_{p_c+k}+m_c)}}}
\end{eqnarray}
Then the decay width $\Gamma(B_c^*\to B_c e^+e^-)$ calculated from the generic expression:
\begin{equation}
\Gamma =\frac{1}{(2\pi)^5}\frac{1}{2M_{B_c^*}}\int\frac{d{\vec k}d{\vec k}_1d{\vec k}_2}{2E_{k_1} 2E_{k_2}} 
\delta ^{(4)}(k+k_1+k_2-{\hat O}M_{B_c^*})
{\bar \sum}_{S_V,\delta}|{\cal M}_{fi}|^2
\end{equation}
is obtained in terms of hadronic $H_{ij}$ and leptonic $L^{ij}$ tensor as
\begin{equation}
\Gamma (B_c^* \to B_c e^+ e^-) = \frac{4\alpha^2_{em}}{(2\pi)^3}\int d^3k{\bar \sum_{S_V,\delta}}H_{ij}L^{ij}
\end{equation}
when,
\begin{equation}
L^{ij}=\int\frac{d{\vec k}_1d{\vec k}_2}{2E_{k_1} 2E_{k_2}} 
\delta ^{(4)}(k+k_1+k_2-{\hat O}M_{B_c^*})\;
Tr[({\not k_2} +m_2)\gamma^i({\not k_1}-m_1)\gamma^j]/(k_1+k_2)^4
\end{equation}
Evaluating trace and adopting standard technique of integration via conversion of three momentum integral to four momentum integral, $L^{ij}$ is simplified to
\begin{equation}
L^{ij}=\frac{2\pi}{3}\frac{\delta^{ij}}{(M_{B_c^*} - {E_k})}
\end{equation}
Due to $\delta^{ij}$ in the expression for $L^{ij}$, $H_{ij}$ is reduced to $H_{ii}$. Note that summing over polarization index and spin states and averaging over $B^*_c$ spin states, one gets ${\bar \sum_{S_V,\delta}} |({\vec \epsilon}\times {\vec k})_i|^2=\frac{2}{3}|{\vec k}|^2$
which leads to the contribution of the hadronic tensor $H_{ii}$ in terms of transition form factor $F_{B_c^*B_c}(q^2)$ as
\begin{equation}
{\bar \sum_{S_V,\delta}}H_{ii}=\frac{|{\vec k}|^2}{3}
|F_{B_c^*B_c}(q^2)|^2 
\end{equation}
Now casting the leptonic and hadronic tensor each as function of $q^2$ and finally integrating out $q^2$ in the kinematic range: $(2m_e)^2\leq q^2\leq(m_{B^*_c}-m_{B_c})^2$, the decay width is obtained in the form
\begin{equation}
\Gamma (B^*_c\to B_ce^+e^-)=\frac{2\alpha^2_{em}}{9\pi M_{B^*_c}}\int_{(2m_e)^2}^{(M_{B^*_c}-M_{B_c})^2}dq^2
\frac{E_k(E^2_k-M^2_{B_c})^{3/2}}{(M_{B^*_c}-E_k)^2}
|F_{B_c^*B_c}(q^2)|^2
\end{equation}
where the energy of $B_c$ is 
$$E_k=\frac{M^2_{B_c^*}-M^2_{B_c}-q^2}{2M_{B_c^*}}$$
In view of recent progress in experimental probe for possible detection of orbitally exited states of $B_c$ and $B^*_c$, we also evaluate $V\to Pe^+e^-$ type transitions: $B^*_c(2s)\to B_c(2s)e^+e^-$, $B^*_c(2s)\to B_ce^+e^-$; $B^*_c(3s)\to B_c(3s)e^+e^-$,
$B^*_c(3s)\to B_c(2s)e^+e^-$, $B^*_c(3s)\to B_ce^+e^-$, and $P\to Ve^+e^-$ type transitions: $B_c(2s)\to B^*_ce^+e^-$, $B_c(3s)\to B^*_c(2s)e^+e^-$,
$B_c(3s)\to B^*_c(1s)e^+e^-$. For $P\to Ve^+e^-$ type transitions the form factor $F_{PV}(q^2)$ can be calculated in the RIQ model as is done above for $F_{VP}(q^2)$ describing $B_c^* \to B_c e^+ e^-$ involving ground states of the participating mesons. The corresponding decay width expression can be obtained in the form:
\begin{equation}
\Gamma [B_c(ns)\to B_c^*(n^{\prime}s)]=\frac{2\alpha^2_{em}}{3\pi M_{B_c(ns)}}\int_{(2m_e)^2}^{(M_{B_c(ns)}-M_{B^*_c(n^{\prime}s)})^2}dq^2
\frac{E_k(E^2_k-M^2_{B^*_c(n^{\prime}s)})^{3/2}}
{(M_{B_c(ns)}-E_k)^2}
|F_{B_cB_c^*}(q^2)|^2
\end{equation}
where $n>n^{\prime}$.
In principle one could extend same analysis to the decay processes involving higher orbital excited states with $n\geq 4$ and P-wave states of the $B_c$- family. But because their production rates are negligibly small and experimental measurements are much more difficult, we do not include those transitions in the present analysis.
\section{Numerical results and discussion}
For numerical analysis of $B_c^* \to B_c e^+ e^-$ involving $B_c^*$ and $B_c$ meson in their ground states, we take relevant quark masses $m_q$,   corresponding binding energy  $E_q$ and potential parameters(a,$V_0$)  which have already been fixed \cite{A35} in the RIQ model by fitting the data of heavy flavored mesons including $B_c$. Using the same set of input parameters, a wide ranging hadronic phenomena \cite{A32, A33, A34, A35, A36, A37, A38, A39, A40, A41} have been described in earlier applications of this model. Accordingly we take 
\begin{eqnarray}
(a, V_0)&\equiv & (0.017166\;{GeV}^3, -0.1375\;GeV)\nonumber\\
(m_b, m_c, E_b, E_c)&\equiv & (4.77659, 1.49276, 4.76633, 1.57951)\;GeV
\end{eqnarray}
Since $B_c^*(1s)$ has not yet been observed, we take our predicted meson masses; $M_{B_c}=6.2642$ GeV and $M_{B_c^*}=6.3078$ GeV \cite {A38} obtained through hyperfine mass splitting in the model.  Our predicted value of $M_{B_c}$ is close to the central value $\sim$ 6.2751GeV of its observed one \cite{A43}.
However for binding energies of constituent  quarks in higher excited states, we solve the cubic equation representing respective bound state condition and obtain
\begin{eqnarray} 
( E_b ; E_c ) = ( 5.05366 ; 1.97016 )  GeV \nonumber\\
( E_b ; E_c ) = ( 5.21703 ; 2.22479 )  GeV 
\end{eqnarray}
for 2s- and 3s states respectively.
With the quark binding energies (23) and other input parameters as in (22), the mass splitting yields $M_{B_c^*}(2s) = 6.78521$ GeV and $M_{B_c^*}(3s) = 6.88501$ GeV. The mass of $B_c(2s)$ so predicted runs short of 57 MeV from the observed value of $6842\pm 4\pm 5$ MeV \cite {A15}. The difficulty encountered here is to make sure all the meson states to have their respective correct masses. This is indeed a problem common to all potential models. Just as in all other model descriptions, we too cannot expect to get precise meson masses for all states with same set of input parameters. So we adjust the potential parameter $V_0$ to a new value $\sim -0.01545$  GeV \cite{A34} as is done by  T.Wang {\it{et al.}} in their work based on the instantaneous approximated Bethe-Salpeter approach \cite {A25}. In doing so we obtain the mass of $B_c(2s)$ equal to its observed value. 
With $V_0 = -0.01545$ GeV and input parameters (22,23), the masses of 
$B_c$, $B_c^*$ meson in 2s- and 3s states are predicted, respectively, as
\begin{eqnarray}
\left (M_{B_c^*}(2s); M_{B_c}(2s)\right) = ( 6910.3 ; 6841.9 ) MeV\nonumber\\
\left (M_{B_c^*}(3s) ; M_{B_c}(3s)\right) = (7259.5; 7135.6) MeV
\end{eqnarray} 
Using appropriate wave packets for participating mesons in hadronic part  and simplifying hadronic and leptonic part separately we calculate the S-matrix element (8-11). Then the invariant transition matrix element ${\cal M}_{fi}$ (12-14) are calculated from which we finally extract the model expression for transition form factor (19).

We then study the $q^2$- dependence of $F_{B_c^*B_c}(q^2)$ and $F_{B_cB_c^*}(q^2)$ for different decay modes in respective kinematic ranges; which are depicted in Fig. (2-3). 
\begin{figure}[!htb]
	\begin{subfigure}{0.48\textwidth}
		\includegraphics[width=\linewidth]{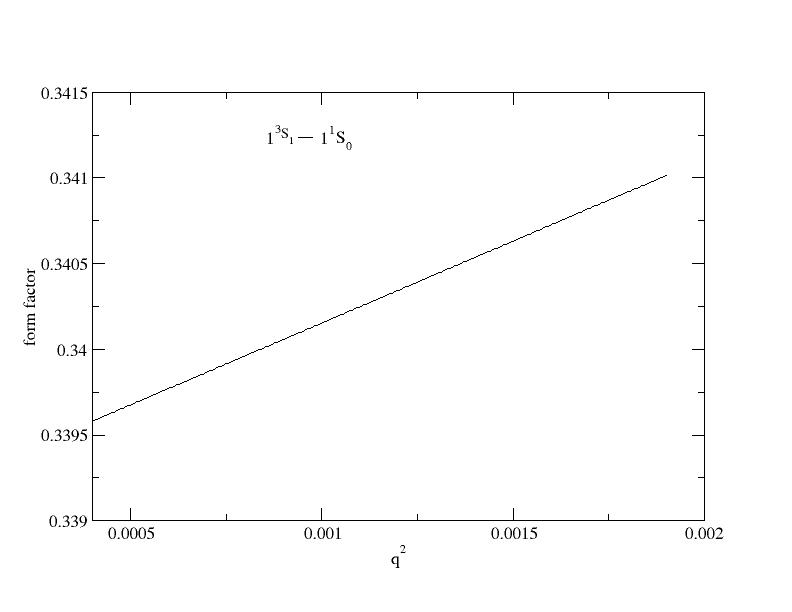}
		\caption{$B^*_c(1s)\to B_c(1s)$} 
	\end{subfigure}
	%\hspace*{\fill} % separation between the subfigures
	\begin{subfigure}{0.48\textwidth}
		\includegraphics[width=\linewidth]{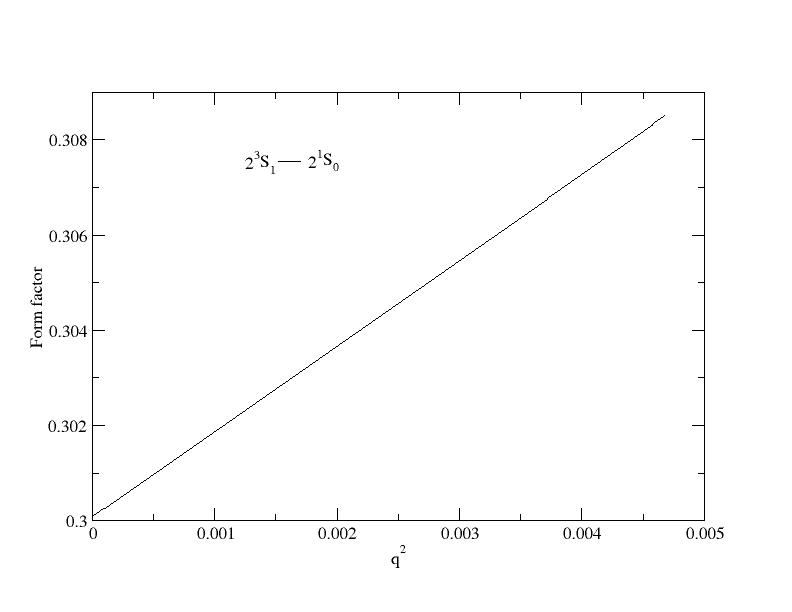}
		\caption{$B^*_c(2s)\to B_c(2s)$} 
	\end{subfigure}
	%\hspace*{\fill} % separation between the subfigures
	\begin{subfigure}{0.48\textwidth}
		\includegraphics[width=\linewidth]{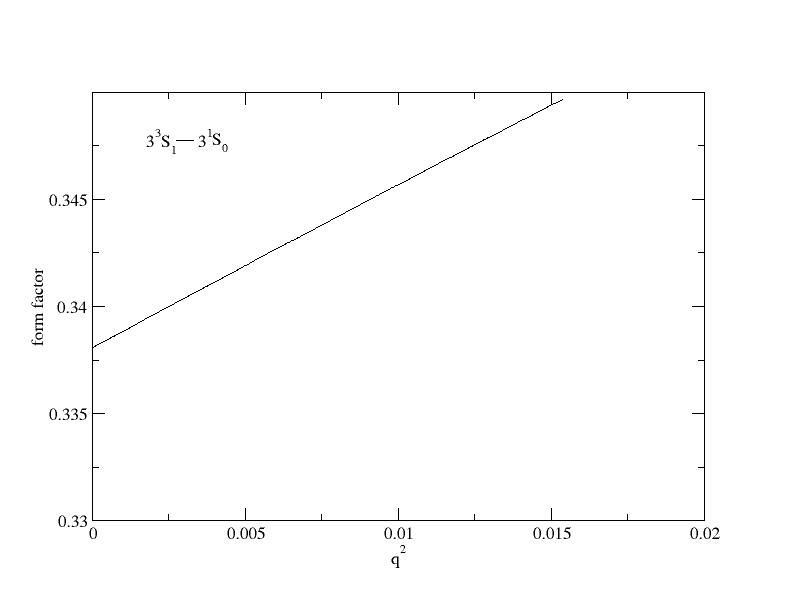}
		\caption{$B^*_c(3s)\to B_c(3s)$} 
	\end{subfigure}
\begin{subfigure}{0.48\textwidth}
	\includegraphics[width=\linewidth]{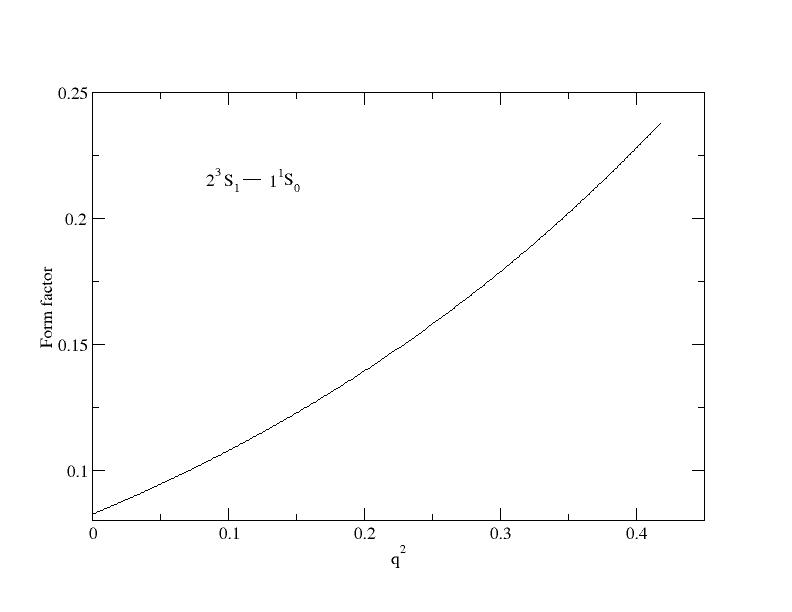}
	\caption{$B^*_c(2s)\to B_c(1s)$} 
\end{subfigure}
\begin{subfigure}{0.48\textwidth}
	\includegraphics[width=\linewidth]{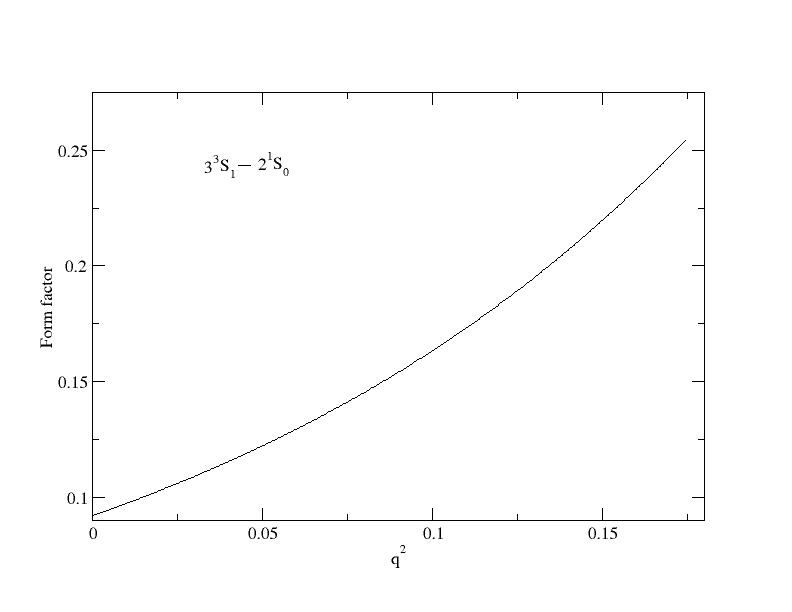}
	\caption{$B^*_c(3s)\to B_c(2s)$} 
\end{subfigure}
\begin{subfigure}{0.48\textwidth}
	\includegraphics[width=\linewidth]{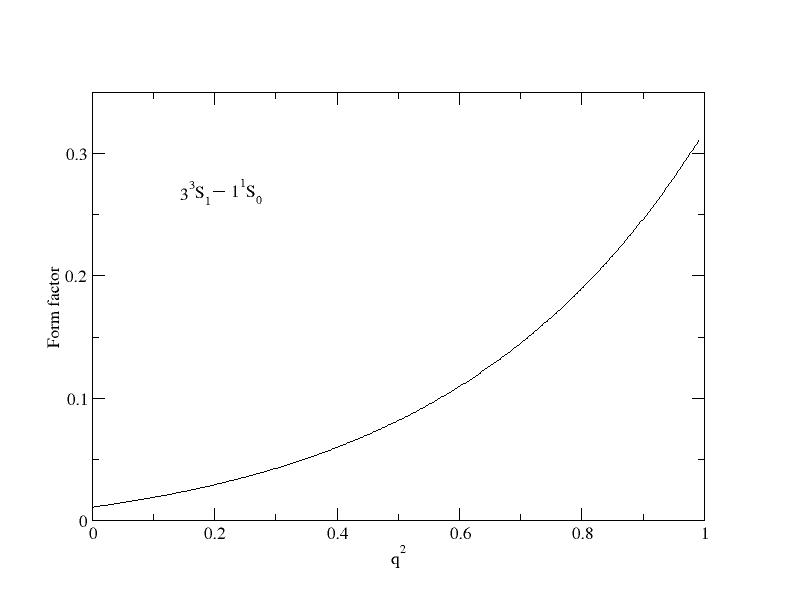}
	\caption{$B^*_c(3s)\to B_c(1s)$} 
\end{subfigure}
		\caption{The $q^2$ dependence of form factor of $B^*_c\to B_c$} 
\end{figure}
For transitions $B_c^*(ns)\to B_c(n^{\prime}s)e^+e^-$ where the mass splitting is marginal the transition form factors are found to increase almost linearly with $q^2$. In other transitions: $B_c^*(ns)\to B_c(n^{\prime}s)e^+e^-$ and  $B_c(ns)\to B^*_c(n^{\prime}s)e^+e^-$ with quantum number $n>n^{\prime}$, where mass difference between participating mesons is comparatively large, $q^2$ dependence of the form factors are found to be parabolic. This is contrary to the predictions of model calculation based on Bethe-Salpeter framework \cite{A17}, where the form factors are found almost constant in respective kinematic range for which they consider $F_{B^*_cB_c}(q^2)=F_{B^*_cB_c}(q^2_{min})$ for their calculation accuracy. However, in present work we do not take resort to such approximation and instead use the calculated form factors as such with their $q^2$ - dependence in respective kinematic range to evaluate decay widths.
\begin{figure}[!htb]
	\begin{subfigure}{0.48\textwidth}
		\includegraphics[width=\linewidth]{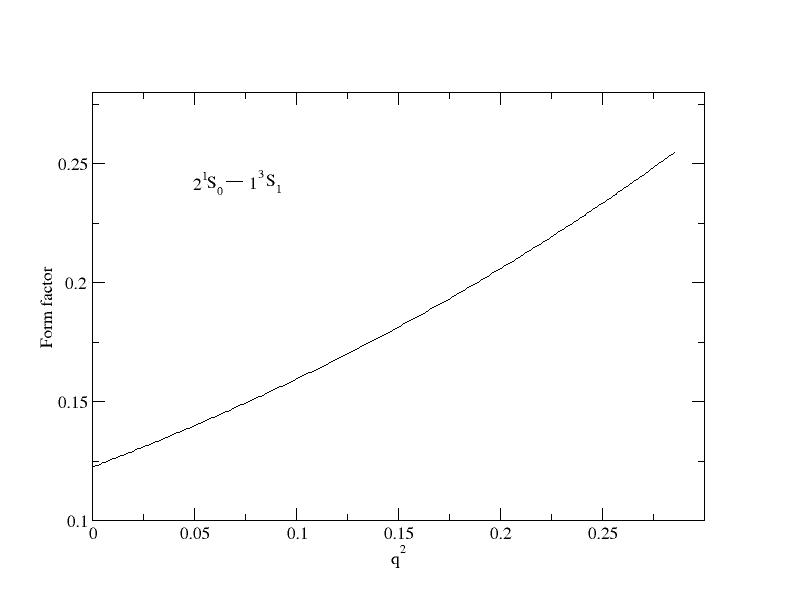}
		\caption{$B_c(2s)\to B^*_c(1s)$} 
	\end{subfigure}
	%\hspace*{\fill} % separation between the subfigures
	\begin{subfigure}{0.48\textwidth}
		\includegraphics[width=\linewidth]{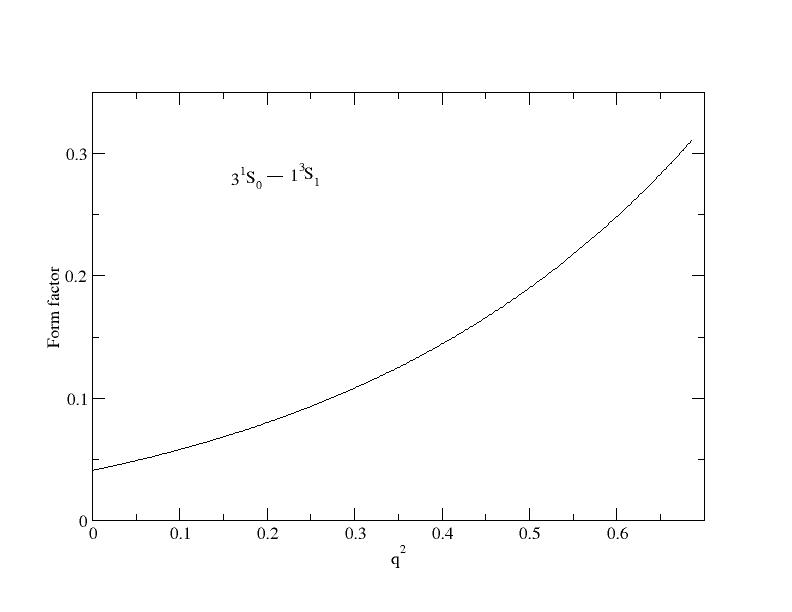}
		\caption{$B_c(3s)\to B^*_c(1s)$} 
	\end{subfigure}
	%\hspace*{\fill} % separation between the subfigures
	\begin{subfigure}{0.48\textwidth}
		\includegraphics[width=\linewidth]{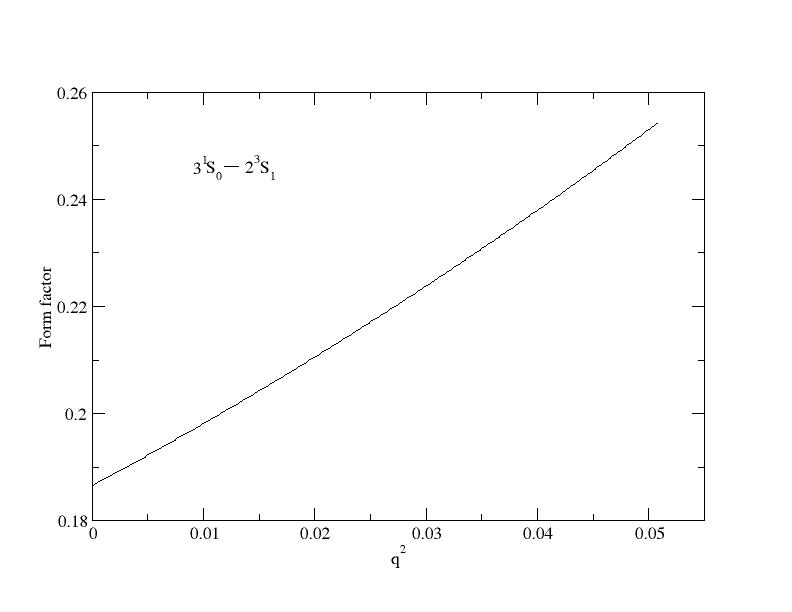}
		\caption{$B_c(3s)\to B^*_c(2s)$} 
	\end{subfigure}
		\caption{The $q^2$ dependence of form factor of $B_c\to B^*_c$} 
\end{figure}

\newpage
\begin{table}[ht]
	\renewcommand{\arraystretch}{1}
	\setlength\tabcolsep{9pt}
	\begin{center}
		\caption{{\bf Comparison of theoretical prediction on decay rates (in KeV) for several electromagnetic decay modes.}}
		\begin{tabular}{|c|c|c|}
			\hline
			\hline
			Transitions & Present work & \cite{A17}\\
			\hline
			$1^{3}S_1\to 1^{1}S_0$ & $0.7112\times10^{-5}$ & $8.64\times10^{-5}$\\
			$2^{3}S_1\to 2^{1}S_0$ & $0.2168\times10^{-4}$ & - \\
			$3^{3}S_1\to 3^{1}S_0$ & $0.1621\times10^{-3}$ &-\\
			$2^{2}S_1\to 1^{1}S_0$ &$0.2452\times10^{-3}$ &$1.59\times10^{-3}$\\
			$3^{3}S_1\to 2^{1}S_0$ &$0.0824\times10^{-3}$ &-\\
			$3^{3}S_1\to 1^{1}S_0$ &$2.3569\times10^{-3}$ &$2.11\times10^{-3}$\\
			$2^{1}S_0\to 1^{3}S_1$ &$0.7297\times10^{-3}$  &$1.65\times10^{-3}$\\
			$3^{1}S_0\to 2^{3}S_1$ &$0.1035\times10^{-3}$ &$1.41\times10^{-3}$\\
			$3^{1}S_0\to 1^{3}S_1$ &$9.4391\times10^{-3}$  &$0.42\times10^{-3}$\\
			\hline
		\end{tabular}
	\end{center}
\end{table}

Then substituting the model expressions  for $F_{B^*_cB_c}(q^2)$ and $F_{B_cB^*_c}(q^2)$ in Eq. (20) and (21), respectively, we evaluate decay widths for transitions involving mesons in 1s, 2s  and 3s states. The predicted decay rates are listed in Table I. 

The transitions: $B_c^*(ns)\to B_c(ns)e^+e^-$ are known as allowed transitions, whereas $B_c^*(ns)\to B_c(n^{\prime}s)e^+e^-$ and $B_c(ns)\to B^*_c(n^{\prime}s)e^+e^-$ together are known as hindered transitions. In the latter type of transitions $\it n$ is greater than $n^{\prime}$. In this work we have analyzed both the allowed and hindered transitions. In the field theoretic description of any decay process the relativistic effects are implicitly incorporated into the analysis by  invoking precise spin-spin interaction while extracting the wave function in the model framework and reproducing hyperfine mass splitting between vector mesons and their pseudoscalar counterparts. In the present work relativistic recoil effect on the antiquark $\bar c$ which is not so heavy  compared to the quark {\it b} is found to be significant. This along with the interaction potential  $U(r)$ taken in equally mixed scalar-vector harmonic form yields the results as shown in Table I. Our predictions on decay widths for both allowed and hindered ones are compatible with the results of model calculations \cite{A17} based on Bethe-Salpeter approach. Although there is an order of magnitude agreement between our predictions and that of \cite{A17}, we find some quantitative disagreement.The future experiments would tell which model is more suitable in describing these decays. The theoretical approaches to describe this decay mode would provide clue for experimental determination of unmeasured meson masses in $B_c$ family. Fortunately, the experiments at LHC and $Z^0$ factory are likely to detect the ground state of $B_c^*$ and other excited states of $B_c$ meson in near future. 

\section {Summary and Conclusion}
We study electromagnetic decays: 
$B_c^*(ns)\to B_c(ns)e^+e^-$, $B_c^*(ns)\to B_c(n^{\prime}s)e^+e^-$ and $B_c(ns)\to B_c^*(n^{\prime}s)e^+e^-$ with the quantum number $n>n^{\prime}$ in the framework of relativistic independent quark model  based on the interaction potential in equally mixed scalar-vector harmonic form. We obtain model expression of  quark and antiquark momentum probability amplitude $G_b(p_b)$ and $\tilde G_c(p_c)$ by taking momentum projections of respective quark orbitals  derived in this model after solving Dirac equation. With an effective momentum profile function considered as 
$${\cal G}_{B_c}({\vec p_b},{\vec p_{\bar c}})=\sqrt{G_b(\vec p_b){\tilde G}_{\bar c}(\vec p_{\bar c})},$$ we construct appropriate wave packets that represent participating meson states at definite momentum and spin and then calculate transition matrix element from which the transition form factors are extracted. For numerical analysis we consider input parameters such as quark mass  $m_q$  and corresponding  binding energy $E_q$  and model parameters (a, $V_0$) which have already been fixed earlier by fitting with heavy flavor data in order to describe the decay process involving ground states. For determining the mass of $B_c$ and $B_c^*$ in their orbitally excited states (2s and 3s), we first calculate binding energies of constituent quarks by solving cubic equations that represent the bound state condition for respective constituent quarks. Then we fine-tune the potential parameter  $V_0$ to a new value $\sim -0.01545$ GeV while retaining the quark masses and model parameters as those used for 1s states and reproduce hyperfine splitting to get psuedoscalar $B_c$(2s)  mass equal to its observed value. However, the transitions involving $B_c$ and $B_c^*$ mesons in 3s states we had to take the same set of input parameters as used for hyperfine splitting of mesons in 2s states as both the $B_c$(3s) and $B_c^*$(3s) states have not yet been observed. With these two sets of input parameters: - one for transition involving 1s state and others  involving 2s and 3s states of $B_c$ and $B_c^*$,  we  obtain numerically the transition form factor  for each $q^2$ value in respective kinematic range. 

Then we study $q^2$ dependence of transition form factor $F_{B^*_cB_c}(q^2)$ and $F_{B_cB^*_c}(q^2)$ for energetically possible transitions of the type $V\to Pe^+e^-$ as well as $P\to Ve^+e^-$ involving ground and orbitally excited states (2s and 3s-states) of $B_c$ family members. We find that for allowed transitions:$B_c^*(ns)\to B_c(ns)e^+e^-$, where the mass splitting is marginal, the transition form factors increase linearly with $q^2$ in the kinetic range of $(2m_e)^2\leq q^2\leq (m_{B_c^*(ns)}-m_{B_c(ns)})^2$ . However for hindered transitions of the type $V\to Pe^+e^-$: $B_c^*(ns)\to B_c(n^{\prime}s)e^+e^-$ and $P\to Ve^+e^-$type: $B_c(ns)\to B_c^*(n^{\prime}s)e^+e^-$ with $n>n^{\prime}$, where mass difference between parent and daughter mesons is comparably large, $q^2$ dependence of relevant transition form factors are found parabolic. Our predictions here are contrary to that obtained in the model calculations based on Bethe-Salpeter approximation. They find the transition form factor almost constant in the entire kinematic range and hence consider $F_{B^*_cB_c}(q^2)=F_{B^*_cB_c}(q^2_{min})$ only for their calculation accuracy. We then substitute the model expression for relevant transition form-factor into the decay width expression, and then integrate out $q^2$ in respective kinematic range and evaluate decay widths for allowed and hindered transitions. In the RIQ model formalism the relativistic effect is incorporated into the analysis by invoking precise spin-spin interaction while extracting the wave function and reproducing mass splitting between vector meson and its pseudoscalar counterpart.  On scrutiny we find relativistic recoil on the antiquark $\bar c$ which is not so heavy compared to the heavy quark $b$ is found to be more significant. This along with our choice of interaction potential in equally mixed scalar-vector harmonic form lead to our predicted decay widths for energetically possible transitions as shown in Table I, which are found compatible with those obtained in the model calculation based on Bethe-Salpeter approach \cite{A17}. There is an order of magnitude agreement between our results and those of \cite {A17}, though there is some quantitative  disagreement. In the absence of precise data in this sector only the future experiments at LHC and $Z^0$-factory would tell which  model is more suitable to provides realistic description of these transitions. Fortunately the experiments at LHCb and particularly $Z^0$-factory are likely to provide precise data in this sector in near future.  
\appendix
\setcounter{section}{0} 
\section{ CONSTITUENT QUARK ORBITALS AND MOMENTUM PROBABILITY AMPLITUDES}

In RIQ model a meson is picturised as a color-singlet assembly of a quark and an antiquark independently confined by an effective and average flavor independent potential in the form:
$U(r)=\frac{1}{2}(1+\gamma^0)(ar^2+V_0)$ where ($a$, $V_0$) are the potential parameters. It is believed that the zeroth order quark dynamics  generated by the phenomological confining potential $U(r)$ taken in equally mixed scalar-vector harmonic form can provide adequate tree level description of the decay process being analyzed in this work. With the interaction potential $U(r)$ put into the zeroth order quark lagrangian density, the ensuing Dirac equation admits static solution of positive and negative energy as: 
\begin{eqnarray}
\psi^{(+)}_{\xi}(\vec r)\;&=&\;\left(
\begin{array}{c}
\frac{ig_{\xi}(r)}{r} \\
\frac{{\vec \sigma}.{\hat r}f_{\xi}(r)}{r}
\end{array}\;\right)U_{\xi}(\hat r)
\nonumber\\
\psi^{(-)}_{\xi}(\vec r)\;&=&\;\left(
\begin{array}{c}
\frac{i({\vec \sigma}.{\hat r})f_{\xi}(r)}{r}\\
\frac{g_{\xi}(r)}{r}
\end{array}\;\right){\tilde U}_{\xi}(\hat r)
\end{eqnarray}
where, $\xi=(nlj)$ represents a set of Dirac quantum numbers specifying 
the eigen-modes;
$U_{\xi}(\hat r)$ and ${\tilde U}_{\xi}(\hat r)$
are the spin angular parts given by,
\begin{eqnarray}
U_{ljm}(\hat r) &=&\sum_{m_l,m_s}<lm_l\;{1\over{2}}m_s|
jm>Y_l^{m_l}(\hat r)\chi^{m_s}_{\frac{1}{2}}\nonumber\\
{\tilde U}_{ljm}(\hat r)&=&(-1)^{j+m-l}U_{lj-m}(\hat r)
\end{eqnarray}
With the quark binding energy $E_q$ and quark mass $m_q$
written in the form $E_q^{\prime}=(E_q-V_0/2)$,
$m_q^{\prime}=(m_q+V_0/2)$ and $\omega_q=E_q^{\prime}+m_q^{\prime}$, one 
can obtain solutions to the resulting radial equation for 
$g_{\xi}(r)$ and $f_{\xi}(r)$in the form:
\begin{eqnarray}
g_{nl}&=& N_{nl} (\frac{r}{r_{nl}})^{l+l}\exp (-r^2/2r^2_{nl})
L_{n-1}^{l+1/2}(r^2/r^2_{nl})\nonumber\\
f_{nl}&=& N_{nl} (\frac{r}{r_{nl}})^{l}\exp (-r^2/2r^2_{nl})\nonumber\\
&\times &\left[(n+l-\frac{1}{2})L_{n-1}^{l-1/2}(r^2/r^2_{nl})
+nL_n^{l-1/2}(r^2/r^2_{nl})\right ]
\end{eqnarray}
where, $r_{nl}= a\omega_{q}^{-1/4}$ is a state independent length parameter, $N_{nl}$
is an overall normalization constant given by
\begin{equation}
N^2_{nl}=\frac{4\Gamma(n)}{\Gamma(n+l+1/2)}\frac{(\omega_{nl}/r_{nl})}
{(3E_q^{\prime}+m_q^{\prime})}
\end{equation}
and
$L_{n-1}^{l+1/2}(r^2/r_{nl}^2)$ etc. are associated Laguerre polynomials. The radial solutions yields an independent quark bound-state condition in the form of a cubic equation:
\begin{equation}
\sqrt{(\omega_q/a)} (E_q^{\prime}-m_q^{\prime})=(4n+2l-1)
\end{equation}
The solution of the cubic equation provides the zeroth order binding energies of 
the confined quark and antiquark for all possible eigenmodes.

In the relativistic independent particle picture of this model, the constituent quark 
and antiquark are thought to move independently inside the $B_c$-meson bound state 
with momentum $\vec p_b$ and $\vec p_c$, respectively. Their individual momentum probability 
amplitudes are obtained in this model via momentum projection of respective quark orbitals (A1) in following forms:
For ground state mesons:($n=1$,$l=0$)
\begin{eqnarray}
G_b(\vec p_b)&=&{{i\pi {\cal N}_b}\over {2\alpha _b\omega _b}}
\sqrt {{(E_{p_b}+m_b)}\over {E_{p_b}}}(E_{p_b}+E_b)\exp {(-{
		{\vec p}^2\over {4\alpha_b}})}\nonumber\\
{\tilde G}_c(\vec p_c)&=&-{{i\pi {\cal N}_c}\over {2\alpha _c\omega _c}}
\sqrt {{(E_{p_c}+m_c)}\over {E_{p_c}}}(E_{p_c}+E_c)\exp {(-{
		{\vec p}^2\over {4\alpha_c}})}
\end{eqnarray}
For excited meson state:($n=2$, $l=0$)
\begin{eqnarray}
G_b(\vec p_b) ={{i\pi {\cal N}_b}\over {2\alpha _b\omega _b}}
\sqrt {{(E_{p_b}+m_b)}\over {E_{p_b}}}\exp {(-{
		{\vec p}^2\over {4\alpha_b}})}\sqrt{(A^2_b+B^2_b)}e^{i\phi_b}\nonumber\\
{\tilde G}_c(\vec p_c)=-{{i\pi {\cal N}_c}\over {2\alpha _c\omega _c}}
\sqrt {{(E_{p_c}+m_c)}\over {E_{p_c}}}\exp {(-{
		{\vec p}^2\over {4\alpha_c}})}\sqrt{(A^2_c+B^2_c)}e^{i{ \phi_c}}
\end{eqnarray}
where,
\begin{eqnarray}
A_{b,c}&=&\frac{3}{\sqrt{\pi}}(E_{p_{b,c}}-m_{b,c})\sqrt{\frac{\alpha_{b,c}}{p^2_{b,c}}}
\;(3-\frac{p^2_{b,c}}{\alpha_{b,c}})\nonumber\\
B_{b,c}&=&\frac{\omega_{b,c}}{2}(\frac{p^2_{b,c}}{\alpha_{b,c}}-3)
+(E_{p_{b,c}}-m_{b,c})(1+\frac{\alpha_{b,c}}{p^2_{b,c}})
\end{eqnarray}

For the excited meson state ($n=3$, $l=0$)
\begin{eqnarray}
G_b(\vec p_b) ={{i\pi {\cal N}_b}\over {4\alpha _b\omega _b}}
\sqrt {{(E_{p_b}+m_b)}\over {E_{p_b}}}
\exp {(-{{\vec p}^2\over {4\alpha_b}})}\sqrt{(A^2_b+B^2_b)}e^{i\phi_b}\nonumber\\
{\tilde G}_c(\vec p_c) =-{{i\pi {\cal N}_c}\over {4\alpha _c\omega _c}}\sqrt {{(E_{p_c}+m_c)}\over {E_{p_c}}}\exp {(-{{\vec p}^2\over {4\alpha_c}})}\sqrt{(A^2_c+B^2_c)}e^{i{\phi_c}}
\end{eqnarray}

where,
\begin{eqnarray}
A_{b,c}&=&\frac{\omega_{b,c}}{2p_{b,c}}\sqrt{\frac{\alpha_{b,c}}{\pi}}
(\frac{5p^4_{b,c}}{\alpha^2_{bc}}-26\frac{{p^2_{b,c}}}
{{\alpha_{b,c}}}-41)\nonumber\\
B_{b,c}&=&\omega_{b,c}(\frac{p^4_{b,c}}{4\alpha^2_{b,c}}-\frac{5p^2_{b,c}}{2\alpha_{b,c}}+\frac{15}{4})+(E_{p_{b,c}}-m_{b,c})\frac{\alpha_{b,c}}{2p^2_{b,c}}(\frac{p^4_{b,c}}{\alpha^2_{b,c}}-\frac{2p^2_{b,c}}{\alpha_{b,c}}+7)
\end{eqnarray}
For both 2s and 3s states: $$\phi_{b,c}=\tan^{-1}\frac{B_{b,c}}{A_{b,c}}$$ with respective $A_{b,c}$ and $B_{b,c}$

The binding energies of the constituent quark and antiquark for ground and orbitally excited $B_c$ and $B_c^*$ states can also be obtained by solving respective cubic equations with $n=1,2,3$ and $l=0$ representing appropriate bound-state conditions by putting the quantum number $n=1,2,3$ and $l=0$.

\begin{acknowledgements}
One of the authors Sonali Patnaik acknowledges the library and computational  facilities provided by the authorities of Siksha 'O' Anusandhan Deemed to be University, Bhubaneswar- 751 030, India to carry out the present work.
\end{acknowledgements}

\end{document}